# Extraordinary Photoluminescence and Strong Temperature/Angle-Dependent Raman Responses in Few-Layer Phosphorene

Shuang Zhang,[1,†] Jiong Yang,[1,†] Renjing Xu,[1] Fan Wang,[2] Weifeng Li,[3] Muhammad Ghufran,[1] Yong-Wei Zhang,[3] Zongfu Yu,[4] Gang Zhang,[3] Qinghua Qin,[1] and Yuerui Lu[1*]

[1]Research School of Engineering, College of Engineering and Computer Science, The Australian National University, Canberra, ACT, 0200, Australia

[2]Department of Electronic Materials Engineering, Research School of Physics and Engineering, The Australian National University, Canberra, ACT, 0200, Australia

[3]Institute of High Performance Computing, A*STAR, 138632, Singapore

[4]Department of Electrical and Computer Engineering, University of Wisconsin, Madison, Wisconsin 53706, USA

[†] These authors contributed equally to this work

[*] To whom correspondence should be addressed. Email: yuerui.lu@anu.edu.au

**ABSTRACT**

**Phosphorene is a new family member of two-dimensional materials. We observed strong and highly layer-dependent photoluminescence in few-layer phosphorene (2 to 5 layers). The results confirmed the theoretical prediction that few-layer phosphorene has a direct and layer-sensitive band gap. We also demonstrated that few-layer phosphorene is more sensitive to temperature modulation than graphene and $MoS_2$ in Raman scattering. The anisotropic Raman response in few-layer phosphorene has enabled us to use an optical method to quickly determine the crystalline orientation without tunneling electron microscope (TEM) or scanning tunneling microscope (STM). Our results provide much needed experimental information about the band structures and exciton nature in few-layer phosphorene.**

Two-dimensional (2D) layered materials, including semi-metallic graphene,[1-7] semiconducting transition metal dichalcogenides (TMDs) [8-11] and insulating hexagonal boron nitride (hBN),[12, 13] have been heavily investigated in past decade. Compared with the gapless graphene, most recently investigated TMD semiconductor $MoS_2$ has energy gap in the range of 1.3 eV (bulk) to 1.8 eV (monolayer). $MoS_2$, an indirect band gap material in its bulk form, becomes a direct band gap semiconductor when thinned to a monolayer, enabling significantly enhanced photoluminescence in monolayer $MoS_2$.[10, 14-16] Black phosphorous (termed as phosphorene) has become a new class of 2D layered material, with predicted layer-dependent band gap ranging from 0.3 eV (bulk) to 1.5 eV (monolayer).[17-22] Particularly, few-layer phosphorene with narrow band gaps ranging from mid-infrared to near-infrared wavelengths can fill the space between the gapless graphene and the comparably large gap TMD semiconductors[22-24]. The predicted direct band gap nature in few-layer phosphorene will also enable high-performance optoelectronic devices, compared with the indirect band gap behavior in most few-layer TMD semiconductors. However, so far there has been very little experimental data to confirm the theoretical prediction in few-layer phosphorene.

**RESULTS AND DISCUSSION**

Here, we report the extraordinarily strong and highly layer-dependent photoluminescence (PL) in few-layer phosphorene, which confirms the predicted direct and layer-dependent band gap nature in few-layer phosphorene. Surprisingly, the PL intensity increases exponentially with decreasing layer thickness from five to two-layer, while the thinner-layer phosphorene has less volume of materials. Our temperature-dependent Raman spectra measurement showed that few-layer phosphorene is more sensitive to temperature than graphene and $MoS_2$. The strong anisotropic properties in phosphorene allows for the fast determination of its crystalline orientation by simple

Raman microscopy, without the requirements of complicated high-resolution imaging systems, like STM or TEM.

Few-layer phosphorene flakes (Figure 1), were fabricated using mechanical exfoliation techniques onto Si/SiO$_2$ chip substrate (270 nm SiO$_2$), similar to that for graphene and MoS$_2$.[7, 10, 25] The flakes were firstly identified by optical contrast in a microscope. Regions with different colors correspond to phosphorene flakes with different thicknesses. Figure 1a displays the optical microscope image of a typical thin phosphorene sample (2 layers, indicated by "2L") on Si/SiO$_2$ substrate. The layer number identification was confirmed by atomic force microscopy (AFM) image of the same sample (Figure 1b). More microscope and AFM images of the phosphorene flakes of different thicknesses are shown in Figure S1.

Optical properties in few-layer phosphorene were investigated by micro-Raman spectroscopy and PL spectroscopy. All the Raman and PL spectroscopy measurements were carried out in a confocal microscopy setup, which has a 532 nm solid state green laser for excitation and a spatial resolution of sub-1 μm. Our PL system has two liquid nitrogen cooled detectors (CCD and InGaAs), which can detect photons with wavelengths ranging from 200 nm to 1650 nm. All the PL spectra in Figure 2a were measured using the InGaAs detector and all the Raman spectra were measured using the CCD detector. In order to prevent the few-layer phosphorene reacting with the moisture or other possible reactants from the environment, the samples were put into a microscope-compatible chamber with a slow flow of nitrogen gas. The measured Raman peaks (Figure 1c), at 359 cm$^{-1}$, 437 cm$^{-1}$ and 466 cm$^{-1}$, are attributed to the $A_g^1$, $B_{2g}$ and $A_g^2$ phonon modes in the crystalline few-layer phosphorene flakes, which matched well with the observations in bulk black phosphorus.[26]

Our measured PL spectra (Figure 2a) in few-layer phosphorous were highly dependent on numbers of layer (2 to 5 layers). Strong PL peaks at 961 nm, 1268 nm, 1413 nm and 1558 nm were observed in 2, 3, 4 and 5-layered phosphorene, respectively, which are corresponding to energy peaks of 1.29 eV, 0.98 eV, 0.88 eV and 0.80 eV, respectively. The measured PL peaks are attributed to the nature of excitons, which represent lower bounds on the fundamental band gap values in few-layer phosphorene, respectively. More PL results on various few-layer (2-5L) phosphorene samples are listed in Table S1 in the Supporting Information. Quite consistent PL center wavelength and peak energy were observed from the samples that have the same layer number. The energy position of the measured PL peak increases rapidly as the layer number decreases. This indirectly confirms our theoretical calculation results that the band gap of few-layer phosphorene increases rapidly as the layer number decreases (Figure 2b), due to the quantum confinement effect. In addition, our measured PL energy peaks from 2-5L phosphorene align well with the reported one from monolayer phosphorene (1.45 eV),[17] which matches the theoretical prediction.

Normally, few-layer phosphorene degrades in air within hours.[27] In our experiments, in order to achieve accurate PL measurements, few-layer phosphorene flakes were quickly identified by optical contrast under optical microscope. Then the sample was rapidly put into a microscope-compatible chamber with a slow flow of nitrogen gas for sample protection. The sample exposure time in air for optical microscope imaging was minimized to be less than around 15 min. After the PL or Raman characterizations, a quick AFM scanning (less than 30 min) in ambient atmosphere at room temperature was used to confirm the layer number of the phosphorene flake. The total time of optical microscope imaging and AFM in ambient air was minimized to be less than 45 min, within which the few-layer phosphorene samples would not degrade too much. Also, we checked the effectiveness of this nitrogen chamber for preventing the possible sample degradation. We measured the optical microscope images and PL spectra of one 5L phosphorene sample, when the sample was just loaded into the nitrogen chamber and after 16 hours. By comparison, the measured PL spectra do not change too much (Figure S2).

We calculated the band structures for 1-5 L phosphorene (Figure S3), using DFT calculations with the implementation of Vienna *ab initio* simulation package (VASP).[28] Hybrid functionals (HSE06)[29] were adopted, together with the projector augmented wave method. A vacuum space of at least 20 Å was kept to avoid mirror interactions. The HSE06 approach is well-known to have an inaccurate description of the dispersion force and thus a poor estimation of the interlayer distance. Here all the structures were relaxed by using the optimized Becke88 van der Waals (optB88-vdW) functional.[29] The first Brillouin zone was sampled with a 10×8×1 Monkhorst-Pack k-meshes for the structure relaxation, and a kinetic energy cut off of 500 eV was adopted. The lattice constant was relaxed until the change of total energy is less than 0.01 meV and all the forces on each atom are less than 0.01 eV/Å, which are sufficient to obtain relaxed structures.

Few-layer phosphorene has only several atoms in thickness, but the PL intensities from 2-5 layered phosphorene were all stronger than that from the thick (hundreds of microns) Si substrate (Figure 2a). This is because of the direct band gap nature in few-layer phosphorene (Figure S3) and the indirect band gap in silicon. Compared with the indirect band gap behaviour in few-layer $MoS_2$,[10] this universal direct band gap nature in few-layer phosphorene grants them great advantages for various optoelectronic applications.

Surprisingly, the PL peak intensity increases dramatically when the layer number decreases, in spite of the reduced amount of material. As the way used in $MoS_2$,[10] we can evaluate the luminescence quantum efficiency $\eta_{PL}$ in few-layer phosphorene by comparing the PL peak intensity $I_{PL}$ that is normalized by layer number. This normalized PL peak intensity (Figure 2c) indirectly reflects the intrinsic luminescence quantum efficiency $\eta_{PL}$.[10] This indicates that the intrinsic luminescence quantum efficiency is highly dependent on layer number in few-layer phosphorene. As shown in Figure 2c, the intrinsic luminescence quantum efficiency in 2L phosphorene is more than one order of magnitude higher than that in 5L phosphorene.

The strong PL in few-layer phosphorene arises from direct electronic transitions with high radiative recombination rate. The internal luminescence quantum efficiency from such direct electronic transitions in few-layer phosphorene can be approximated by $\eta_{PL} \approx k_{rad}/(k_{rad} + k_{defect} + k_{relax})$, where $k_{rad}$, $k_{defect}$, and $k_{relax}$ are, respectively, rates of radiative recombination, defect trapping, and electron relaxation within the conduction and valence bands, similar to that in $MoS_2$.[10] Both $k_{rad}$ and $k_{relax}$ are highly related to the density of states and the band structures.[30] This strong layer-dependent internal quantum efficiency can be understood from the layer-dependent band structures in few-layer phosphorene (Figure S3). As layer number increases from monolayer to five-layer, more and more band valleys and band maxima show up both within the conduction and valence bands, respectively. Those emerging valleys (maxima) in the conduction (valence) band will significantly change the density of states distribution for electrons (holes). Especially, those states, provided by those valleys or maxima at the off-$\Gamma$ points, will highly enhance the relaxation rates within the conduction or valence bands, leading to lower internal quantum efficiency in 5L phosphorene compared to that in 2L. Our experimental data points towards the rich PL physics in few-layer phosphorene, which calls for further in-depth theoretical and experimental analysis to fully understand carrier dynamics.

Raman spectroscopy is another powerful tool to characterize the structural and electronic properties of few-layer phosphorene.[31, 32] Particularly, temperature-dependent Raman study for few-layer phosphorene is important to further understand the fine structure and properties of the material, such as atomic bonds, thermal expansion, and thermal conductivity. The temperature dependence of the Raman spectra, measured at temperature ranging from 20 °C to -160 °C in a 5L phosphorene, is shown in Figure 3. The decreasing temperature leads to the blue shift of the Raman phonon modes, $A_g^1$, $B_{2g}$ and $A_g^2$. The measured temperature dependence of the Raman mode frequency shift in few-layer phosphorene can be characterized by a linear equation: $\omega = \omega_0 + \chi T$, where $\omega_0$ is the mode frequency at zero K and $\chi$ is the first-order temperature coefficient. The measured $\chi$ values for modes $A_g^1$, $B_{2g}$ and $A_g^2$ in the 5L phosphorene are -0.023 cm$^{-1}$/°C, -0.018 cm$^{-1}$/°C and -0.023 cm$^{-1}$/°C, respectively (Figure 3b). The change of the Raman shift with temperature is determined by the anharmonic terms in the lattice potential energy, which is related to the anharmonic potential constants, the phonon occupation number, and the thermal expansion of the crystal.[33] These measured $\chi$ values in 5L phosphorene are all larger in absolute value than those from both bi-layer graphene (-0.0154 cm$^{-1}$/°C for G peak) [31] and few-layer MoS$_2$ (-0.0123 cm$^{-1}$/°C and -0.0132 cm$^{-1}$/°C for $A_{1g}$ and $E_{2g}^1$ modes respectively)[32]. This indicates that the phonon frequencies in few-layer phosphorene are more sensitive to temperature modulation than those in graphene and MoS$_2$, which could be due to the superior mechanical flexibility of phosphorene originated from its unique puckered crystal structure.[34]

Compared with other 2D materials, phosphorene shows strong anisotropic properties,[17, 35] allowing for various unique applications in optoelectronics. Normally, TEM is used to analyze the crystalline orientation of crystals. But the high-energy electrons in TEM is likely to introduce lots of defects in phosphorene, which can expedite the reaction of phosphorene with moisture and other possible reactants from the environment.[27] Thus, it is important to develop optical means to characterize the crystalline properties. It has been recently shown that second-harmonic generation is used to determine the crystalline orientation of MoS$_2$.[36] Both polarized reflection spectroscopy[24] and polarized IR spectroscopy[23] are capable to provide simple and non-destructive optical ways to determine the crystalline orientation of phosphorene. Here we further report an alternative and simple optical method based on anisotropic Raman response to determine the crystalline orientation of phosphorene.

We performed linearly polarized Raman measurements to quickly determine the crystalline orientation of a phosphorene flake (15L) (Figure 4). The normally incident laser was in the z direction. The polarization angle θ is relative to the zero degree reference, which can be arbitrarily selected at the beginning. We observed that the Raman intensities of $B_{2g}$ and $A_g^2$ modes are significantly dependent on the polarization angle (Figure 4a). The angle-dependent intensities of $B_{2g}$ and $A_g^2$ modes both show an angle period of 180º and are out-of-phase (Figure 4b). The Raman intensity of $A_g^1$ is less sensitive to the polarization angle (Figure 4b). Our measurement results can be perfectly explained by the vibration directions of these three Raman modes in crystalline black phosphorous.[26] Figure 4c shows a schematic of the vibration direction of the phosphorus atoms in the different Raman modes.[26] In $A_g^2$, $B_{2g}$ and $A_g^1$ vibrational modes, the phosphorus atoms oscillate along the x (zigzag), y (armchair) and z (out-of-plane) directions, respectively. When the laser polarization is parallel to x direction (y direction), the intensity of the $A_g^2$ mode will reach the maximum (minimum) value, meanwhile, the intensity of the $B_{2g}$ mode will reach the minimum (maximum) value. Our measurement result (Figure 4b) highly matched this theoretical prediction.

Based on this polarization-dependent Raman measurement, the crystalline orientation of this 15L phosphorene flake was quickly determined, as indicated in Figure 4d. This technique provides a fast and precise determination of the crystalline orientation, without the need of complicated and high-resolution imaging systems, like STM or TEM. Here, we just considered the incident laser polarization and the oscillation direction of the atom vibration. In order to more accurately interpret the polarization dependence of $A_g^2$, $B_{2g}$ and $A_g^1$ Raman modes, more factors might need to be considered, such as polarization of the scattering light. Considering this, the polarization dependence of $A_g^1$ is probably intrinsic.

**CONCLUSIONS**

In conclusion, we observed strong and layer-sensitive PL in 2-5 layered phosphorene. The energy gaps, determined from the PL spectra, are highly dependent on the numbers of layers, which matches well with the theoretical simulation. This PL measurement provides very useful information to study the exciton nature and the electronic structures in few-layer phoshoene. Our temperature-dependent Raman measurements demonstrated that few-layer phosphorene is more sensitive to temperature modulation than graphene and MoS$_2$. Also, the highly angle-dependent Raman spectra from a phosphorene flake enabled us to fast determine the crystalline orientation of the flake without TEM or STM.

**EXPERIMENTAL METHODS**

The thin-layer phosphorene were transferred onto SiO$_2$/Si substrate (275 nm thermal SiO$_2$) by mechanical exfoliation from bulk black phosphorus single crystal (from Smart-elements), using GEL film (Gel-Pak). Few-layer phosphorene flakes were quickly identified by optical contrast under optical microscope. Then the sample was rapidly put into a microscope-compatible chamber with a slow flow of protection nitrogen gas, in order to slow down the reaction of phosphorene with moisture and other possible reactants from the environment. The sample exposure time in air for optical microscope imaging was minimized to be less than around 15 min. All PL, Raman and polarization measurements were conducted with a T64000 micro-Raman system equipped with both CCD and InGaAs detectors. For low temperature Raman measurements, a Linkam THMS 600 liquid nitrogen low temperature stage was added onto the micro-Raman system. And for polarization measurements, we inserted an angle-variable polaroid into the optical path of the micro-Raman system and shifted the polaroid 6 degrees every time and then proceeded 30 times to get a full period of polarization results. To avoid laser-induced sample damage, all Raman and PL spectra were recorded at low power levels P ~ 100 μW. Raman measurements used short integration times of ~10 sec and PL used longer integration times of ~ 60 sec. The nitrogen chamber and low power laser can effectively protect the sample. After the PL or Raman characterizations, a quick AFM scanning (less than 30 min) was used to confirm the layer number of the phosphorene flake, which was carried out in ambient atmosphere at room temperature with a Bruker MultiMode III AFM.

## Acknowledgements

We would like to thank Professor Chennupati Jagadish, Professor Barry Luther-Davies and Professor Vincent Craig from The Australian National University, for their facility support. We acknowledge financial support from ANU PhD student scholarship, Wisconsin Alumni Research Foundation, Australian Research Council and ANU Major Equipment Committee.

## Competing financial interests

The authors declare that they have no competing financial interests.

**Supporting Information Available:** Details of theoretical calculation methods, more experimental PL measurement results and characterization of the sample protection by nitrogen flow chamber, are shown in Supporting Information. This material is available free of charge *via* the Internet at http://pubs.acs.org.

**FIGURE CAPTIONS**

**Figure 1. Images and characterization of exfoliated phosphorene.** (a) Microscope image of 2L phosphorene. (b) AFM image of 2L phosphorene, with region indicated in the dash line box in (A). (c) Raman spectrum of 2L phosphorene. (d) Schematic plot of phosphorene layer structure.

**Figure 2. Photoluminescence (PL) spectra of thin-layer phosphorene.** (a) PL spectra of 2L, 3L, 4L and 5L phosphorene. Note: the tiny oscillation on the PL curve of 3L is due to the limitation error of the InGaAs detector. (b) Energy gap of 2L, 3L, 4L and 5L phosphorene from experimental PL spectra and theoretical simulation. (c) Layer dependence of PL peak intensity that is normalized by layer number.

**Figure 3. Low-temperature Raman spectra of 5L phosphorene.** (a) Raman spectra of a 5L phosphorene at temperature ranging from 20 °C to -160 °C. (b) Temperature dependence of $A_g^1$, $B_{2g}$ and $A_g^2$ Raman peak positions.

**Figure 4. Phosphorene crystalline orientation determination by polarization Raman spectra.** (a) Raman spectra of a 15L phosphorene under different polarization angles. (b) Polarization dependence of $A_g^1$, $B_{2g}$, $A_g^2$ modes in a 15L phosphorene and the Raman peaks in silicon. (c) Schematic plot showing the vibration directions of $A_g^1$, $B_{2g}$ and $A_g^2$ Raman modes. (d) Crystalline orientation of the 15L phosphorene flake, determined by angle-dependent Raman measurement.

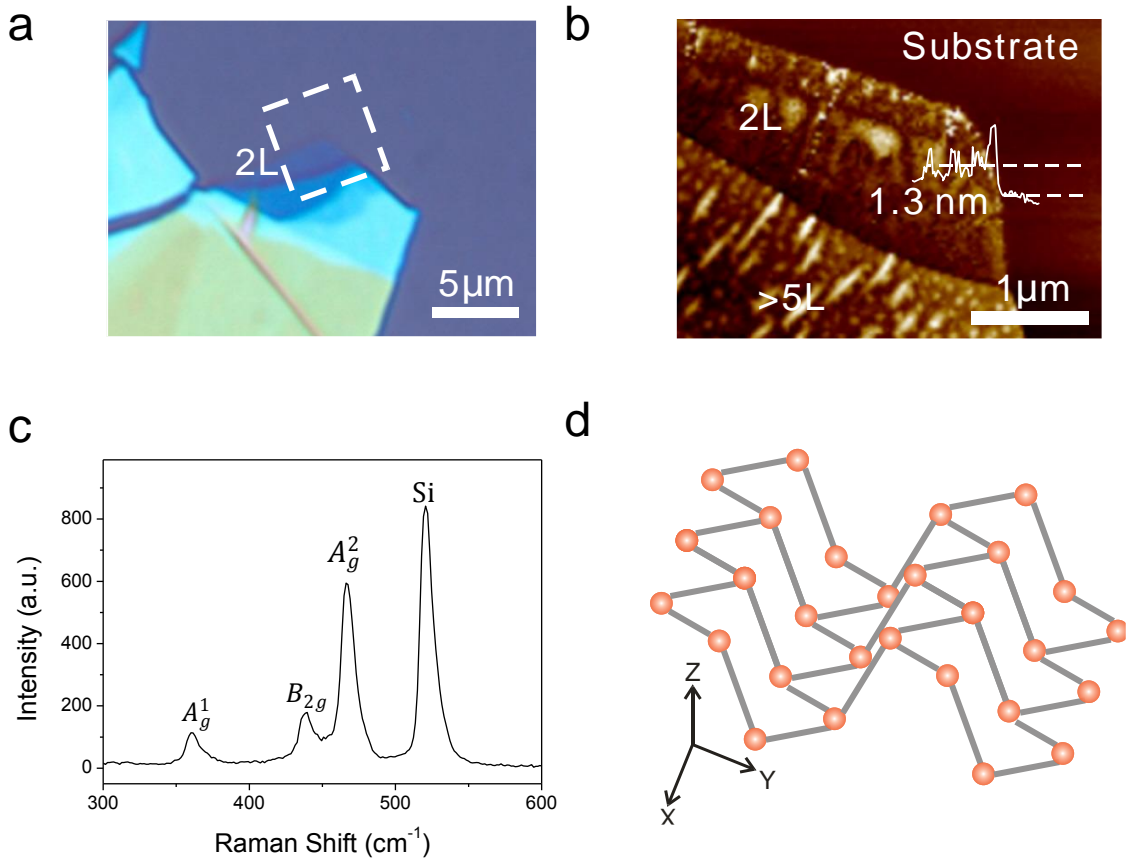

Figure 1

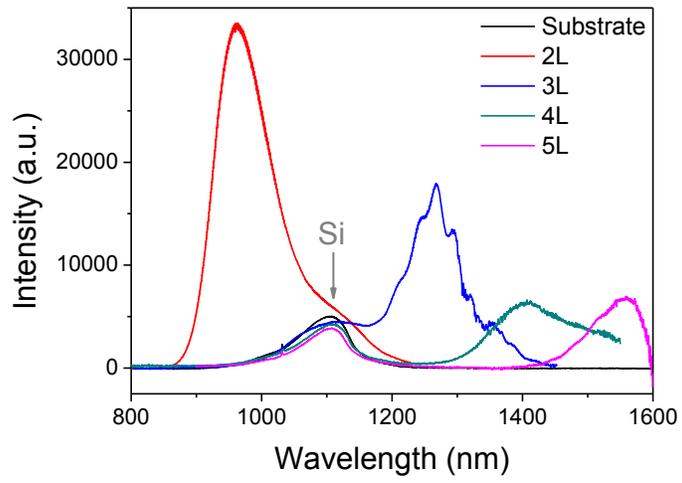

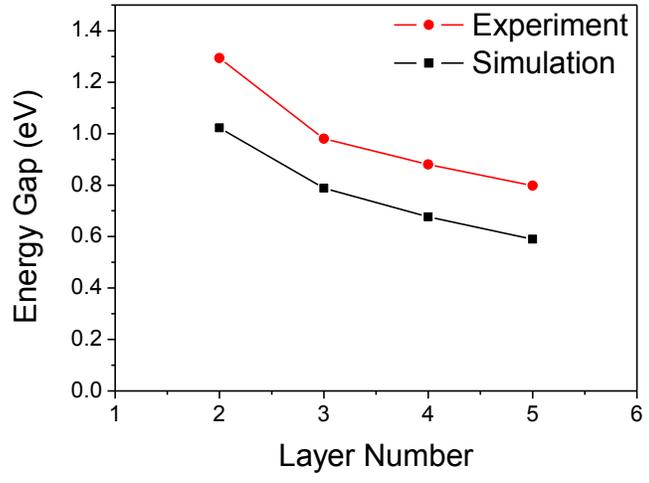

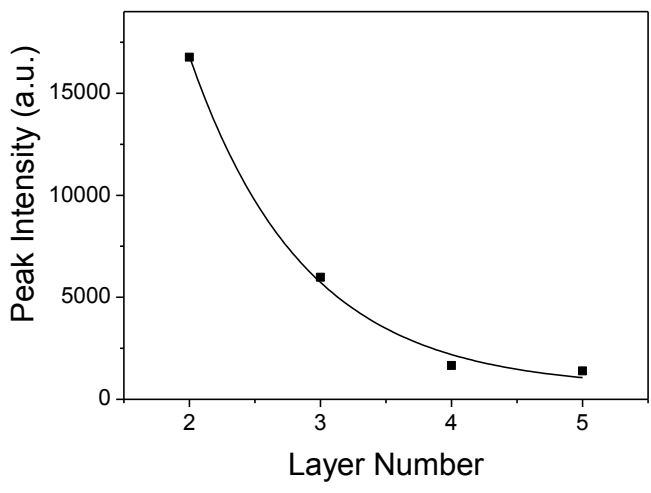

**Figure 2**

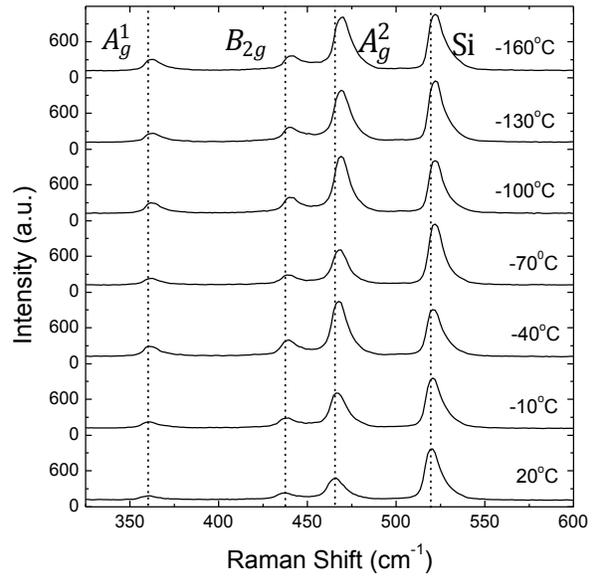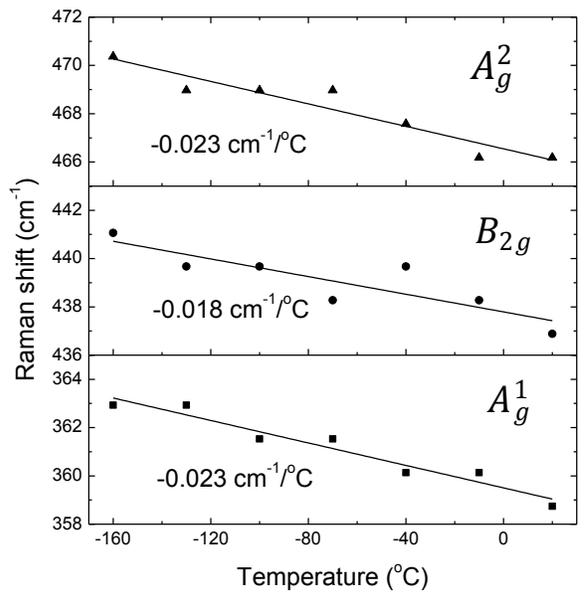

**Figure 3**

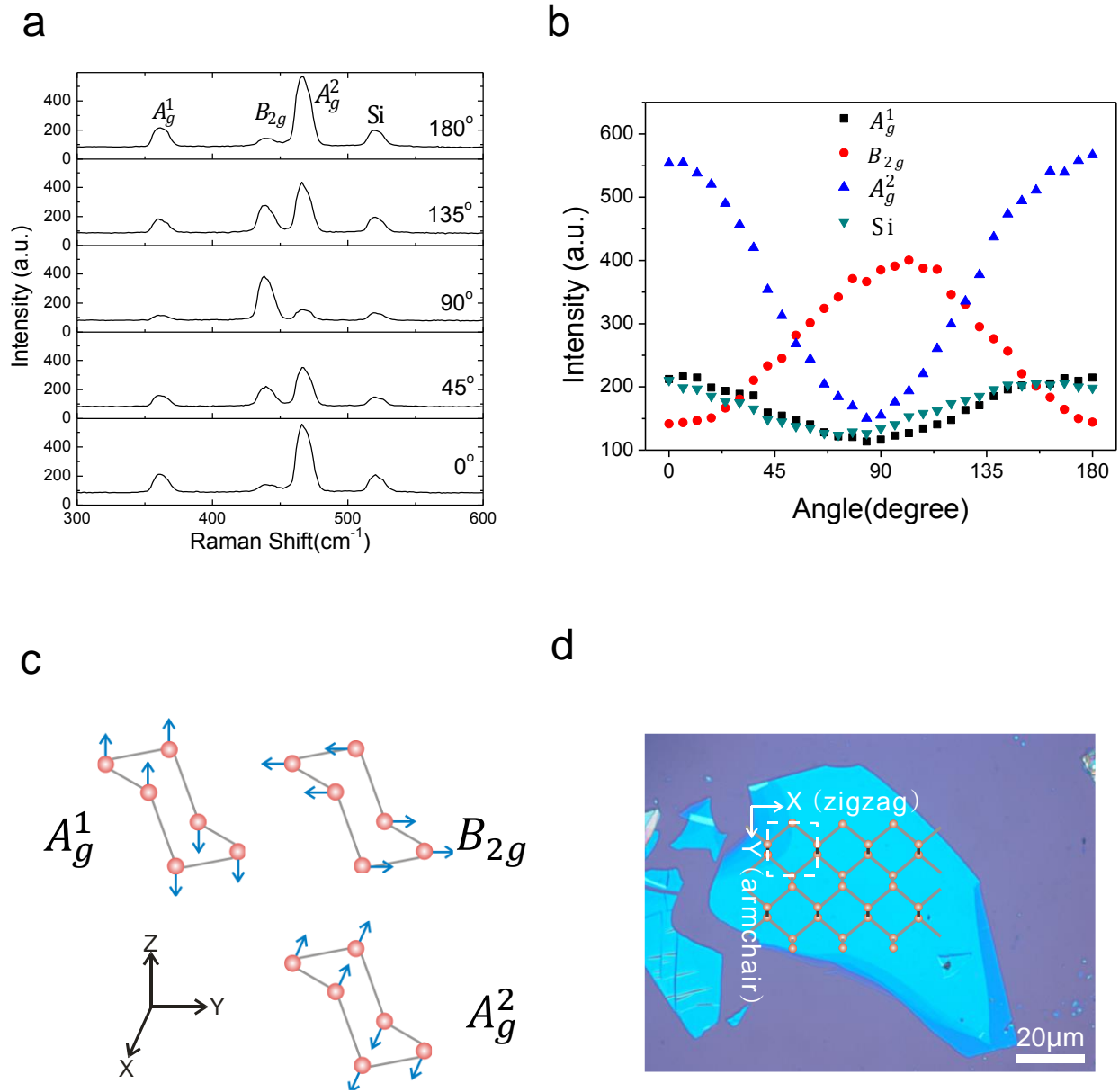

**Figure 4**

**ToC graphic**

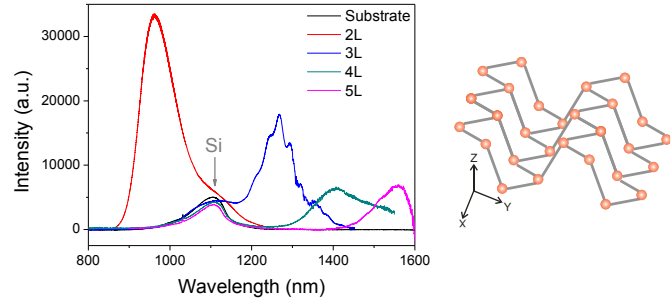

# Supporting Information for

# Extraordinary Photoluminescence and Strong Temperature/Angle-Dependent Raman Responses in Few-Layer Phosphorene

Shuang Zhang,[1,†] Jiong Yang,[1,†] Renjing Xu,[1] Fan Wang,[2] Weifeng Li,[3] Muhammad Ghufran,[1] Yong-Wei Zhang,[3] Zongfu Yu,[4] Gang Zhang,[3] Qinghua Qin,[1] and Yuerui Lu[1*]

[1]Research School of Engineering, College of Engineering and Computer Science, The Australian National University, Canberra, ACT, 0200, Australia

[2]Department of Electronic Materials Engineering, Research School of Physics and Engineering, The Australian National University, Canberra, ACT, 0200, Australia

[3]Institute of High Performance Computing, A*STAR, 138632, Singapore

[4]Department of Electrical and Computer Engineering, University of Wisconsin, Madison, Wisconsin 53706, USA

[†] These authors contributed equally to this work

**\*** To whom correspondence should be addressed. Email: yuerui.lu@anu.edu.au

1. PL measurement data from various few-layer phosphorene samples

The PL spectra of multiple few-layer phosphorene samples were characterized (Table S1). Quite consistent PL center wavelength and peak energy were observed from the samples that have the same layer number.

**Table S1: PL measurement data from various few-layer phosphorene samples.**

| Samples | No. of layers | PL center wavelength (nm) | Peak energy (eV) |
| --- | --- | --- | --- |
| Sample-1 | 2L | 961 | 1.29 |
| Sample-2 | 2L | 965 | 1.29 |
| Sample-3 | 3L | 1268 | 0.98 |
| Sample-4 | 3L | 1264 | 0.98 |
| Sample-5 | 4L | 1413 | 0.88 |
| Sample-6 | 4L | 1441 | 0.86 |
| Sample-7 | 5L | 1558 | 0.80 |
| Sample-8 | 5L | 1555 | 0.80 |
| Sample-9 | 5L | 1557 | 0.80 |

2. **Theoretical calculation methods**

In the calculations, the number of few-layer phosphorene was considered from 1L to 5L. All the structural optimization and band structure calculations were performed using DFT calculations implemented in the plane wave code Vienna *ab initio* simulation package (VASP).[1,2] Hybrid functionals (HSE06)[3] were adopted, together with the projector augmented wave method. A vacuum space of at least 20 Å was kept to avoid mirror interactions. The HSE06 approach is well-known to have an inaccurate description of the dispersion force and thus a poor estimation of the interlayer distance. Here all the structures were relaxed by using the optimized Becke88 van der

Waals (optB88-vdW) functional.[4] The first Brillouin zone was sampled with a 10×8×1 Monkhorst-Pack k-meshes for the structure relaxation, and a kinetic energy cut off of 500 eV was adopted. The lattice constant was relaxed until the change of total energy is less than 0.01 meV and all the forces on each atom are less than 0.01eV/Å, which are sufficient to obtain relaxed structures. The relaxed lattice constant for the 1L phosphorene is a = 3.34 Å and b = 4.57 Å, in good agreement with previous study.[5] Figure S2 (b)-(f) shows the band structures for 1L to 5L phosphorene calculated using HSE06.

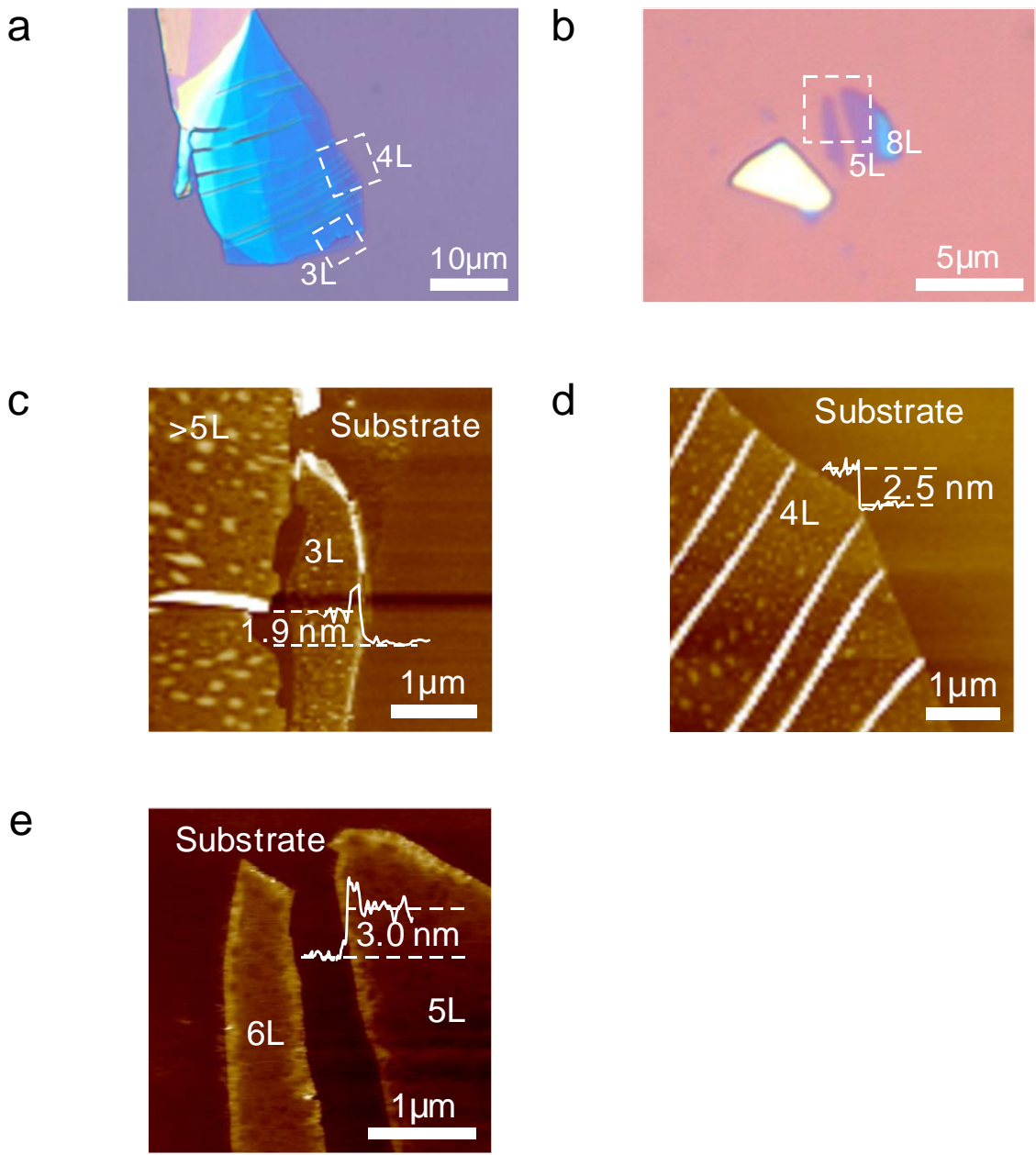

**Figure S1. Microscope and AFM images of 3L, 4L and 5L phosphorene.** (a) Microscope image of 3L and 4L phosphorene. (b) Microscope image of 5L phosphorene. (c)-(e) show the AFM images of 3L, 4L and 5L phosphorene, respectively.

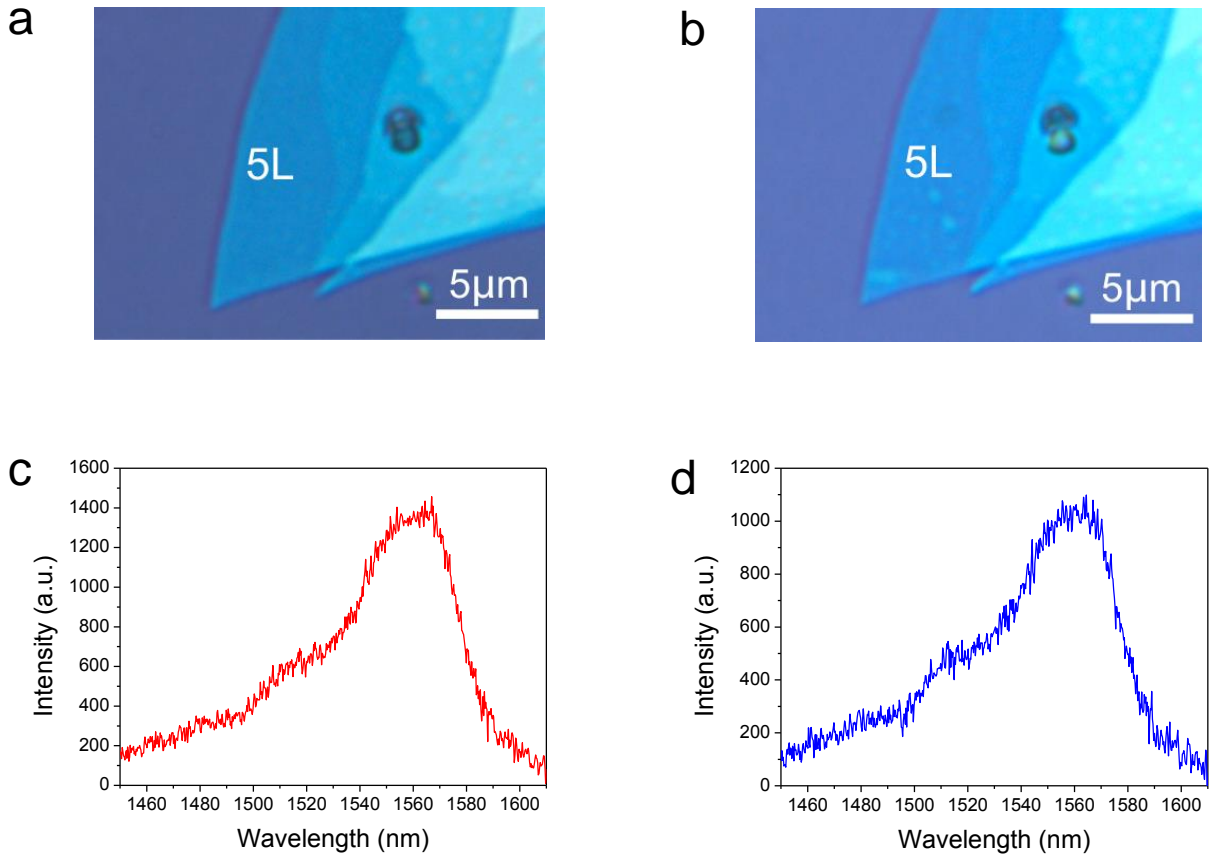

**Figure S2. Phosphorene performance before and after nitrogen chamber protection.** (a)-(b) Optical microscope images of a 5L phosphorene in nitrogen chamber protection when it was just loaded in (a) and after 16 hours (b). (c)-(d) Measured PL spectra of the 5L phosphorene in nitrogen chamber protection when it was just loaded in (c) and after 16 hours (d).

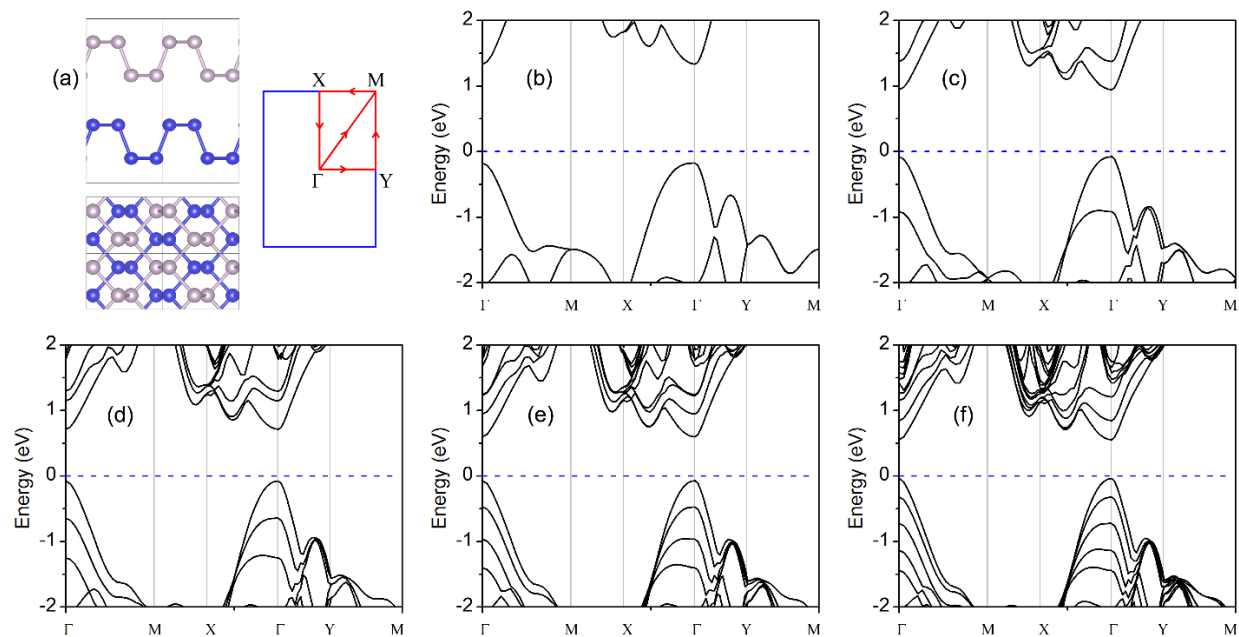

**Figure S3. Simulated band structures for phosphorene (1-5 layers) based on DFT calculation.**

(a) Schematic plot showing the lattice structure of phosphorene. (b)-(f) shows the band structures for 1L to 5L phosphorene calculated using HSE06.